\documentstyle[12pt,twoside,psfig]{article}
\begin{document}
\begin{center}{\Large\bf  A divergence free parametrization of deceleration parameter for scalar field dark energy}
\\[15mm]
Abdulla Al Mamon\footnote{E-mail : abdullaalmamon.rs@visva-bharati.ac.in}~and
Sudipta Das\footnote{E-mail:  sudipta.das@visva-bharati.ac.in}\\
{\em Department of Physics, Visva-Bharati,\\
Santiniketan- 731235, ~India.}\\
[15mm]
\end{center}
\vspace{0.5cm}
\vspace{0.5cm}
\pagestyle{myheadings}
\newcommand{\be}{\begin{equation}}
\newcommand{\ee}{\end{equation}}
\newcommand{\bea}{\begin{eqnarray}}
\newcommand{\eea}{\end{eqnarray}}
\newcommand{\bc}{\begin{center}}
\newcommand{\ec}{\end{center}}
%%%%%%%%%%%%%%%%%%%%%%%%%%%%%%%%%%%%%%%%%%%%%%%%%%%%%%%%%%%%%%%%%%%%%%%
%%%%%%%%%%%%%%%%%%%%%%%%%%%%%%%%%%%%%%%%%%%%%%%%%%%%%%%%%%%%%%%%%%%%%%%%%%%%%%
\begin{abstract}
In this paper, we have considered a spatially flat FRW universe filled with pressureless matter and dark energy. We have considered a phenomenological parametrization of the deceleration parameter $q(z)$ and from this we have reconstructed the equation of state for dark energy $\omega_{\phi}(z)$. This divergence free parametrization of the deceleration parameter is inspired from one of the most popular parametrization of the dark energy equation of state given by Barboza and Alcaniz \cite{ba}. Using the combination of datasets (SN Ia $+$ Hubble $+$ BAO/CMB), we have constrained the transition redshift $z_{t}$ (at which the universe switches from a decelerating to an accelerating phase) and have found the best fit value of $z_{t}$. We have also compared the reconstructed results of $q(z)$ and $\omega_{\phi}(z)$ and have found that the results are compatible with a $\Lambda$CDM universe if we consider SN Ia $+$ Hubble data but inclusion of BAO/CMB data makes $q(z)$ and $\omega_{\phi}(z)$ incompatible with $\Lambda$CDM model. The potential term for the present toy model is found to be functionally similar to a Higgs potential. 
\end{abstract}
%%%%%%%%%%%%%%%%%%%%%%%%%%%%%%%%%%%%%%%%%%%%%%%%%%%%%%%%%%%%%%%%%%%%%%%%%%%%%%%%%%
\section{Introduction}
The discovery of the late-time cosmic acceleration \cite{Riess, Perlmutter} opened up a new field of research in modern cosmology. A number of theoretical models have been constructed to explain this accelerated phenomenon. Most of them are based either on some modification of the Einstein-Hilbert action \cite{Vollick, Nojiri} or the existence of new kind of exotic fields in nature, dubbed as ``dark energy" (DE). In this paper, we will focus on the second aspect and consider DE as the driving agent for the current accelerated expansion of the universe which is considered as a hypothetical energy component with a large negative pressure. In the last decade numerous DE models have been explored to account for this phenomenon (for review, see references \cite{de1,de2,de3,de4,de5}). In spite of those efforts, however, the true nature of dark energy still remains a mystery. The most popular and simplest cosmological DE model is the $\Lambda$CDM model, which is in good agreement with the recent observational data.  The $\Lambda$CDM model is obtained by introducing a cosmological constant $\Lambda$ into general relativity, for which the equation of state parameter $\omega_{\Lambda} = -1$. However, it suffers from two major problems, namely, {\it fine tuning} and cosmological {\it coincidence} problems \cite{sw, Steinhardt}. This motivates theorists to develop some alternative models of DE. Scalar field models, viz, quintessence models, are widely used in addressing several issues in cosmology due to their dynamical nature \cite{Chiba,Armendariz-Picon,Kamenshchik,Caldwell,SMC,SN} for which the equation of state parameter evolves dynamically with time contrary to the cosmological constant models. Usually, the potential term for the quintessence field dominates over its kinetic part in order to generate sufficient negative pressure and generates acceleration. A large variety of quintessence potentials have been considered to serve the purpose. But, none of these models have very strong observational evidence. Some excellent reviews on this topic can be found in \cite{de4, sahnipot} and the references therein.\\
%%%%%%%%%%%%%%%%%%%%%%%%%%%%%%%%%%%%%%%%%%%%%%%%%%%%%%%%%%%%%%%%%%%%%%%%%%%%%%%%%%%%%%%%%%%
%%%%%%%%%%%%%%%%%%%%%%%%%%%%%%%%%%%%%%%%%%%%%%%%%%%%%%%%%%%%%%%%%%%%%%%%%%%%%%%%%%%%%%%
\par From the theoretical viewpoint, in the absence of DE, the universe should be decelerating as gravity holds matter together. The existence of an early decelerated expansion phase of the universe is also supported by the gravitational instability theory of structure formation and of big bang nucleosynthesis. The deceleration parameter, thus,  must show a signature flip in order to incorporate both the scenarios. In other words, a transition from decelerating phase ($q>0$) to a late-time accelerating phase ($q<0$) is necessary to explain the structure formation as well as the current acceleration measurements. Recently, various studies have been performed to analyze the kinematics of the universe through phenomenological parametrizations of $q(z)$, in a model independent way (see references\cite{dp1,dp2,dp3,dp4,dp5,dp6,dp7,dp8,dp9,dp10,dp11}). However, in most of these cases, the $q$-parametrization is valid for $z<<1$ only, and others diverge at $z\rightarrow -1$ (see references\cite{dp2,dp3,dp7,dp8,dp9,dp10}). Recently a new parametrization for DE equation of state $\omega_{\phi}(z)$ has been considered which diverges neither in the past nor in future \cite{ba,akarsu}, however very few such 
divergenceless parametrizations appear in the literature. The basic characteristics of dynamical evolution, both static and dynamic, can be expressed in terms of the Hubble parameter $H$ and the deceleration parameter $q$ only. Infact these two parameters enable us to construct model-independent kinematics of the cosmological expansion \cite{ylbolotin}. Usually, {\it kinematic approach} is described by a particular metric theory of gravity. It does not depend on the validity of general relativity or any model specific assumptions like the matter-energy content of the universe. Motivated by this facts, in the present work, we propose a simple two-parameter parametric form of $q(z)$ to constrain the evolution behavior of DE such that $q(z)$ does not diverge for the entire range $z\in [-1,\infty ]$ and we study the expansion history of the universe for this divergence-free parametrization. The functional form of $q$-parametrization is similar to the one presented in \cite{ba} for equation of state parameter $\omega_{\phi}$. The properties of this parametrization has been discussed in detail in the next section. Based on this consideration, we have then solved the field equations for this model and have obtained the expressions for different cosmological parameters, such as the equation of state parameter ($\omega_{\phi}(z)$). We have also constrained $q(z)$, $\omega_{\phi}(z)$ and the potential term using the combination of SN Ia, $H(z)$ and BAO/CMB dataset to investigate the various properties of this model.\\
%%%%%%%%%%%%%%%%%%%%%%%%%%%%%%%%%%%%%%%%%%%%%%%%%%%%%%%%%%%%%%%%%%%%%%%%%%%%%%%%
%%%%%%%%%%%%%%%%%%%%%%%%%%%%%%%%%%%%%%%%%%%%%%%%%%%%%%%%%%%%%%%%%%%%%%%%%%%%%%%%%%%%%%%%%  
\par This paper is organised as follows. In section 2, we have described the basic theoretical framework for the quintessence model of a spatially flat FRW universe. We have chosen a particular form of $q(z)$ to solve the field equations and have reconstructed the equation of state parameter $\omega_{\phi}(z)$. In section 3, we have considered observational datasets of SN Ia, $H(z)$ and BAO/CMB and have studied the constraints on the various reconstructed model parameters and have summarized the main results of this analysis in section 4. Finally, in the last section, some conclusions are presented on the basis of the results obtained in this work.
\section{Field equations and Results}
%%%%%%%%%%%%%%%%%%%%%%Basic Framework%%%%%%%%%%%%%%%%%%%%%%
For a universe having space-time curvature $R$ filled with a scalar field $\phi$ having a potential $V(\phi)$ and normal matter, the action is given by
\be\label{action}
S = \int\sqrt{-g} d^{4}x{\left[\frac{R}{2} + \frac{1}{2} \phi^{,\mu}\phi_{,\mu} - V(\phi) + L_{m}\right]}
\ee
where $L_{m}$ is the Lagrangian of the background matter which is considered as pressureless perfect fluid. Here, all quantities are expressed in units of $8{\pi}G=c=1$.\\
\par For a spatially flat FRW space-time with the metric
\be\label{metric}
ds^{2} = dt^{2} - a^{2}(t)[dr^{2} + r^{2}d{\theta}^{2} +r^{2}sin^{2}\theta d{\phi}^{2}]
\ee
the Einstein field equations are obtained as   
\be\label{eq1}
3H^{2} = {\rho}_{m} + \frac{1}{2}{\dot{\phi}}^2 + V(\phi)
\ee
\be\label{eq2}
2{\dot{H}} + 3H^{2} = -\frac{1}{2}{\dot{\phi}}^2 + V(\phi)
\ee
where an overdot indicates differentiation with respect to time $t$ and $\rho{_m}$ represents the energy density of the background matter component of the universe. We choose a spatially flat geometry, which is favoured by the updated results of the cosmic background radiation measurement \cite{Bernardis}. In a FRW background, the energy density $\rho_{\phi}$ and pressure density $p_{\phi}$ of the scalar field $\phi$ are
\be
\rho_{\phi} = \frac{1}{2}{\dot{\phi}}^2 + V(\phi),\hspace{5mm} p_{\phi} = \frac{1}{2}{\dot{\phi}}^2 - V(\phi), \hspace{5mm} \omega_{\phi}=\frac{p_{\phi}}{\rho_{\phi}}
\ee
Here $\omega_{\phi}$ is the equation of state (EoS) parameter of the scalar field. The conservation equation for the scalar field and matter field are 
\be\label{eq3}
{\ddot{\phi}} + 3H{\dot{\phi}} + \frac{dV}{d\phi} = 0
\ee
\be\label{eq4}
{\dot{\rho}}_{m} + 3H{\rho}_{m} = 0
\ee
Among the four equations (equations (\ref{eq1}), (\ref{eq2}), (\ref{eq3}) and (\ref{eq4})), only three are independent equations with four unknown parameters $H$, $\rho_{m}$, $\phi$ and $V(\phi)$. So we still have freedom to choose one parameter to close the above system of equations.\\ 
%%%%%%%%%%%%%%%%%%%%%%%%%%%%%%%%%%%%%%%%%%%%%%%%%%%%%%%%%%%%%%%%
\par The deceleration parameter is defined as
\be\label{qdef}
q = -\frac{{\ddot{a}}}{aH^2} = -(1 + \frac{\dot{H}}{H^2})
\ee
where, $q>0$ corresponds to decelerating phase whereas $q<0$ indicates accelerating phase of the universe. Equation (\ref{qdef}) integrates to yield
\be\label{hq}
H(z)=H_{0}{\rm exp}{\left(\int^{z}_{0}\frac{1+q(z^{\prime})}{1+z^{\prime}}dz^{\prime}\right)}
\ee
where, $H_{0}$ is the Hubble parameter at the present epoch and $z=\frac{1}{a}-1$ is the redshift. Clearly, if the functional form of $q(z)$ is known, then one can obtain information regarding the evolution of the Hubble parameter.\\
In principle, we can parametrize the deceleration parameter as,
\be\label{taylor}
q(z)=\sum_{n=0}q_{n}f_{n}(z)
\ee
where $q_{n}$'s are some real numbers and $f_{n}(z)$ are functions of redshift $z$. This includes constant $q$ model, when $f_{0}(z)=1$ and $f_{n}(z)=0$ ($n\ge 1$). It deserves mention that a large number of parametrizations of $q(z)$ have been considered in literature \cite{dp1,dp2,dp3,dp4,dp5,dp6,dp7,dp8,dp9,dp10,dp11},but the exact functional form of $q(z)$ that will describe the whole evolution history of the universe is still being searched for. 
%%%%%%%%%%%%%%%%%%%%%%%%%%%%%%%%%%%%%%%%%%%%%%%%%%%%%%%%%%%%%%%%
%%%%%%%%%%%%%%%%%%%%%%%%%%%%%%%%%%%%%%%%%%%%%%%%%%%%%%%%%%%%%
\par In this present work, in order to close the system of equations, we have considered a new parametric form of $q(z)$ as
\be\label{qba}
q(z)=q_{0}+q_{1}\frac{z(1+z)}{1+z^{2}}
\ee
In the above parametrization, $q_{0}$ represents the present value of $q(z)$ and the second term represents the variation of the deceleration parameter with respect to $z$. The deceleration parameter reduces to $q(z)=q_{0} + q_{1}$ ,  when $z>>1$ (at high redshift). One can thus obtain a radiation dominated universe by suitably choosing values of $q_{0}$ and $q_{1}$ for $z>>1$ limit. However for small $z$, the model represents dark energy behaviour. It should also be noted that the above parametrization generalizes to the well known linear expansion for $q(z)=q_{0} + q_{1}z$ \cite{dp4}, when $z<<1$ (at low redshift). The advantage of this parametrization is that it provides  finite value of $q$ in the entire range, $z\in [-1,\infty]$ and is thus valid for the entire evolution history of the universe. So, we can use the parametrization for studying the future evolution of the universe also. It is worth noting here that the above parametric form for $q(z)$ was inspired from one of the most popular divergence-free parametrization of the dark energy equation of state \cite{ba}. We have considered the functional form of $q(z)$ containing two terms only in order to make the model simple and elegant. Functional forms of $q(z)$ containing suitable higher order terms may provide more information regarding the evolution history of the universe but will involve complicated algebra and lacks simplicity. But these non-trivial parametrizations always opens up possibilities for future studies regarding the nature of DE. We can now easily constrain the two parameters $q_{0}$ and $q_{1}$ by using the available observational data.\\ %In the following section, the data analysis used for this parametrization is described.\\
\par Inserting equation (\ref{qba}) into the equation (\ref{hq}), we obtained the expression for the Hubble parameter (in terms of $z$) as
\be\label{hz}
H(z)=H_{0}(1+z)^{(1+q_{0})}(1+z^{2})^{\frac{q_{1}}{2}}
\ee
For the sake of completeness, solving (\ref{eq4}), we obtain
\be\label{rhom}
\rho_{m}(z)=\rho_{m0}(1+z)^{3}
\ee
Here $\rho_{m0}$ is an integration constant and represents the present value of the matter field density. From equations (\ref{eq1}), (\ref{hz}) and (\ref{rhom}), we obtained the expression for energy density of the scalar field as
\be\label{rhop}
\rho_{\phi}(z)=3H^{2}_{0}(1+z)^{2(1+q_{0})}(1+z^{2})^{q_{1}} - 3H^{2}_{0}\Omega_{m0}(1+z)^{3}
\ee
Here $\Omega_{m0}=\frac{\rho_{m0}}{3H^{2}_{0}}$ is current value of the density parameter of the matter field.\\
For this model, the expressions for the density parameters of the matter field ($\Omega_{m}$) and the scalar field ($\Omega_{\phi}$) as well as the equation of state parameter ($\omega_{\phi}(z)$) can be easily obtained (in terms of $z$) as
%%%%%%%%%%%%%%%%%%%%%%%%%%%%%%%%%%%%%%%%%%%%%%%%%%%%%%%%%%%%%%%%%%%%%%%%%%%%%%%%%%%%%%%%%%%%%%
\be
\Omega_{m}(z) \equiv \frac{\rho_{m}(z)}{3H^2(z)}=\Omega_{m0}(1+z)^{(1-2q_{0})}(1+z^2)^{-q_{1}}
\ee
%%%%%%%%%%%%%%%%%%%%%%%%%%%%%%%%%%%%%%%%%%%%%%%%%%%%%%%%%%%%%%%%%%%%%%%%%%%%%%%%%
\be
\Omega_{\phi}(z) \equiv \frac{\rho_{\phi}(z)}{3H^2(z)}=1- \Omega_{m}(z)~~~~~~~~~~~~~~~~~~~~~~~~~
\ee
%%%%%%%%%%%%%%%%%%%%%%%%%%%%%%%%%%%%%%%%%%%%%%%%%%%%%%%%%%%%%%%
\be\label{wphi}
\omega_{\phi}(z)=\frac{2{\left[q(z)-\frac{1}{2}\right]}}{3{\left[1-\Omega_{m}(z)\right]}}=\frac{2{\left(q_{0}+q_{1}\frac{z(1+z)}{1+z^{2}}-\frac{1}{2}\right)}}{3-3\Omega_{m0}(1+z)^{(1-2q_{0})}(1+z^{2})^{-q_{1}}}
\ee\\
%%%%%%%%%%%%%%%%%%%%%%%%%%%%%%%%%%%%%%%%%%%%%%%%%%%%%%%%%%%%%%%%%%%%%%%%%%%%%%%%%%%%%%%%
%%%%%%%%%%%%%%%%%%%%%%%%%%%%%%%%%%%%%%%%%%%%%%%%%%%%%%%%%%%%%%%%%%%%%%%%%%%%%%%%%%%%%%%%
From equation (\ref{wphi}), it can be easily seen:\\ \\
1. For $z<<1$, the EoS parameter $\omega_{\phi}(z)$ reduces to
\be\label{canoeqzl1wphi}
\omega_{\phi}(z)=\frac{2q_{0}-1}{3-3\Omega_{m0}} + \frac{2q_{1}z}{3-3\Omega_{m0}}
\ee
which is similar to the linear redshift parametrization of $\omega_{\phi}(z)$ given by $\omega_{\phi}(z)=\omega_{0} + \omega_{1}z$, which has been used for several cosmological analysis (see references \cite{linearw, linearw1}).\\ \\
%%%%%%%%%%%%%%%%%%%%%%%%%%%%%%%%%%%%%%%%%%%%%%%%%%%%%%%%%%%%%%%%%%%%%%%%%%%%%%%%%%%
2. For $z>>1$, we have
\be\label{canoeqzg1wphi}
\omega_{\phi}(z)=-\frac{1-2q_{0}-2q_{1}}{3-3\Omega_{m0} z^{1-2q_{0}-2q_{1}}}
\ee
which is the form of $\omega_{\phi}(z)=-\frac{\omega_{2}}{1+\omega_{3}z^{\beta}}$ (where $\omega_{2}$, $\omega_{3}$ and $\beta$ are arbitrary constants). This parametrization is similar to the parametrization of $\omega_{\phi}(z)$ \cite{ankan} (for $z>>1$) for suitably chosen values of $\omega_{2}$, $\omega_{3}$ and $\beta$.\\ \\
%%%%%%%%%%%%%%%%%%%%%%%%%%%%%%%%%%%%%%%%%%%%%%%%%%%%%%%%%%%%%%%%%%%%%%%%%%%%%%%%%%%%%
%%%%%%%%%%%%%%%%%%%%%%%%%%%%%%%%%%%%%%%%%%%%%%%%%%%%%%%%%%%%%%%%%%%%%%%%%%%%%%%
Obviously, the new parametrization of $\omega_{\phi}(z)$ or $q(z)$ covers a wide range of other models and it also shows a bounded behavior for both high and low redshifts. The simplicity of the functional form of $\omega_{\phi}(z)$, however, makes it very attractive and simple to study.\\
%%%%%%%%%%%%%%%%%%%%%%%%%%%%%%%%%%%%%%%%%%%%%%%%%%%%%%%%%%%%%%%%%%%%%%%%%%%%%%%%%%%%%%%%%%%%%
%%%%%%%%%%%%%%%%%%%%%%%%%%%%%%%%%%%%%%%%%%%%%%%%%%%%%%%%%%%%%%%%%%%%%%%%
\par Now equation (\ref{eq2}) can be re-written as
\be
{\dot{H}}=-\frac{1}{2}({\dot{\phi}}^{2}+\rho_{m})
\ee
Substituting ${\dot{H}}=-(1+z)H\frac{dH}{dz}$ in the above equation, one obtains
\be\label{phidot}
\frac{1}{3H^{2}_{0}}{\dot{\phi}}^{2}=\frac{(1+z)}{3H^{2}_{0}}\frac{dH^{2}}{dz} - \Omega_{m0}(1+z)^{3}
\ee
%%%%%%%%%%%%%%%%%%%%%%%%%%%%%%%%%%%%%%%%%%%%%%%%%%%%%%%%%%%%%%%%%%%%%%%%%%%%%%%%
$\Rightarrow$ $\frac{1}{3H^{2}_{0}}{\dot{\phi}}^{2}=\frac{2}{3}[q_{1}z(1+z)^{3+2q_{0}} (1+z^2)^{q_{1}-1} + (1+q_{0})(1+z)^{2(1+q_{0})}(1+z^2)^{q_{1}}]$
\be\label{pdreco}
-\Omega_{m0}(1+z)^3 
\ee
which immediately gives (by replacing ${\dot{\phi}}=-(1+z)H\frac{d\phi}{dz}$)
\be\label{phiz}
\phi = \phi_{0} + \int^{z}_{0}{\left[\frac{2q_{1}z^{\prime}}{(1+z^{\prime})(1+{z^{\prime}}^2)}+\frac{2(1+q_{0})}{(1+z^{\prime})^2}-\frac{3\Omega_{m0}}{(1+z^{\prime})^{1+2q_{0}}(1+{z^{\prime}}^2)^{q_{1}}}\right]}^{\frac{1}{2}}dz^{\prime}
\ee
where $\phi_{0}$ is an integration constant. Using equations (\ref{eq1}) and (\ref{phidot}), then the effective potential for the quintessence field can be reconstructed as
%%%%%%%%%%%%%%%%%%%%%%%%%%%%%%%%%%%%%%%%%%%%%%%%%%%%%%%%%%%%%%%%%%%%%%%%%%%%%%%%%%%%%%%%%%%
\be
\frac{1}{3H^{2}_{0}}V(z)=\frac{H^{2}}{H^{2}_{0}} - \frac{(1+z)}{6H^{2}_{0}}\frac{dH^{2}}{dz}-\frac{\Omega_{m0}}{2}(1+z)^{3}
\ee
Now, substituting $H$ from equation (\ref{hz}) in the above equation, we obtain\\ \\
$\frac{1}{3H^{2}_{0}}V(z)=(1+z)^{2(1+q_{0})}(1+z^{2})^{q_{1}} -\frac{1}{2}\Omega_{m0}(1+z)^{3}$
\be\label{vz}
-{\frac{(1+z)}{3}{\left[q_{1}z (1+z)^{2(1+q_{0})}(1+z^{2})^{q_{1}-1} + (1+q_{0})(1+z)^{(1+2q_{0})}(1+z^{2})^{q_{1}}     
 \right]}}
\ee
Therefore, we can reconstruct the kinetic term ${\dot{\phi}}^2$ and the potential $V(z)$ (in units of critical density $\rho_{crit}=3H^{2}_{0}$) using equations (\ref{pdreco}) and (\ref{vz}) if the value of $\Omega_{m0}$ is given.\\
%%%%%%%%%%%%%%%%%%%%%%%%%%%%%%%%%%%%%%%%%%%%%%%%%%%%%%%%%%%%%%%%%%%%
Now, we will apply the same limits used for $\omega_{\phi}(z)$ in equation (\ref{wphi}) to study the quintessence potential in more detail.
From equations (\ref{phiz}) and (\ref{vz}), it can be easily seen for the following limiting cases:\\ \\
1. For $z<<1$, the functional forms of ${\dot{\phi}}^{2}$ and $V(z)$ turn out to be
\be
\phi(z)= \phi_{0} + \frac{1}{3q_{1}} {\left[2(1+q_{0}) -3\Omega_{m0} + 2q_{1}z\right]}^{\frac{3}{2}}
\ee
and
\be
V(z) =3H^{2}_{0} \left(\frac{2}{3} -\frac{\Omega_{m0}}{2} -\frac{q_{0}}{3}\right) - 3H^{2}_{0} q_{1}z
\ee
Combining the above two sets of equations we get,
\be
V(\phi)=V_{0}+\eta (\phi - \phi^{*})^{\frac{2}{3}}
\ee
where $V_{0}=H^{2}_{0}(1-\Omega_{m0})$, $\eta=-\frac{H^{2}_{0}}{2} (3q_{1})^{\frac{2}{3}}$ and $\phi^{*}=\phi_{0}-\frac{[2(1+q_{0})-3\Omega_{m0}]^{\frac{3}{2}}}{3q_{1}}$. Note that the above potential is almost similar to the well-known power-law potential for appropriate choices of $V_{0}$, $\eta$ and $\phi^{*}$.\\ \\
2. Similarly, for $z>>1$, we have obtained
\be\label{eqcanovzggovz}
\frac{1}{3H^{2}_{0}}V(z) =  \left(\frac{2}{3}-\frac{q_{0}}{3}-\frac{q_{1}}{3}\right)z^{2(1+q_{0}+q_{1})} -\frac{1}{2}\Omega_{m0}z^{3}
\ee
and 
\be\label{eqcanovzggovz1}
\phi(z)= \phi_{0} + \int_{z}{\left[\frac{2(1+q_{0}+q_{1})}{{z^{\prime}}^2} -\frac{3\Omega_{m0}}{{z^{\prime}}^{(1+2q_{0}+2q_{1})}}\right]}^{\frac{1}{2}}dz^{\prime}
\ee
%%%%%%%%%%%%%%%%%%%%%%%%%%%%%%%%%%%%%%%%%%%%%%%%%%%%%%%%%%%%%%%%%%%%%
From equations (\ref{eqcanovzggovz}) and (\ref{eqcanovzggovz1}), it is not straightforward to obtain the functional form of $V(\phi)$. However, parametric plot of equations (\ref{eqcanovzggovz}) and (\ref{eqcanovzggovz1}) reveals that in the high $z$ limit, the form of $V(\phi)$ is a polynomial in $\phi$ of order four.
\par As seen from equation (\ref{vz}), the functional form of $V(z)$ obtained involves a number of model parameters and thus depends crucially on the values of these parameters. One can obviously choose the model parameters arbitrarily and look at the functional behaviour of the dynamical cosmological parameters obtained. These arbitrarily chosen values can then be confronted with observational data. But here we do the other way around. We first constraint the various model parameters using the available datasets and with the best fit values obtained, we try to reconstruct the functional dependence of $V(z)$ or $V(\phi)$.
%%%%%%%%%%%%%%%%%%%%%%%%%%%%%%%%%%%%%%%%%%%%%%%%%%%%%%%%%%%%%%%%%%%%%%%%%%%%%%%%%%%%%%%%%%%%%
%%%%%%%%%%%%%%%%%%%%%%%%%%%%%%%%%%%%%%%%%%%%%%%%%%%%%%%%%%%%%%%%
\section{Data analysis and method}
We shall fit the theoretical model with the recent observational datasets, namely $H(z)$, SN Ia, BAO and CMB datasets. With the combined datasets, we obtain constraints on the model parameters and then try to obtain the functional form of $V(\phi)$ in the subsequent sections. For the sake of completeness, in this section we describe very briefly the techniques followed for different datasets.
\subsection{Hubble dataset ($H(z)$)}
To constraint cosmological parameter using Hubble dataset, the $\chi^2$ function is defined as
\be
\chi^2_{H} = \sum^{29}_{i=1}\frac{[{h}^{obs}(z_{i}) - {h}^{th}(z_{i})]^2}{\sigma^2_{H}(z_{i})} 
\ee
where ${h} = \frac{H(z)}{H_{0}}$ is the normalized Hubble parameter. In above equation subscript $``obs"$ refers to observational quantities and subscript $``th"$ is for the corresponding theoretical ones. Also, the error for normalized Hubble parameter is given by 
\be
\sigma_{h}=h\sqrt{{\left(\frac{\sigma_{H}}{H}\right)}^{2} + {\left(\frac{\sigma_{H_{0}}}{H_{0}}\right)}^{2}}
\ee
where, $\sigma_{H}$ and $\sigma_{H_{0}}$ are the errors in $H$ and $H_{0}$ respectively. In this work, we have used the 29 data points of Hubble parameter measurements \cite{h1,h2,h3,h4,h5,h6,h7,h8,h9} in the redshift range $0.07\le z\le 2.34$ (see Table 1 of reference \cite{aam}). The current value of the Hubble parameter $H_{0}$ is taken from reference \cite{h0}.
%%%%%%%%%%%%%%%%%%%%%%%%%%%%%%%%%%%%%%%%%%%%%%%%%%%%%%%%%%%%%%%%%%%%%%%%%%%%%%
\subsection{Type Ia Supernova dataset (SN Ia)}
Here we have used the recently released Union2.1 compilation (Suzuki et al. \cite{sn1a}) which contains 580 data points with redshift ranging from 0.015 to 1.414. The $\chi^2$ for SN Ia is defined as (for more details see \cite{sndatamethod})
\be\label{chisquare} 
\chi^2_{SN} = A - \frac{B^2}{C}
\ee
where $A$, $B$ and $C$ are defined as follows
\bea
A = \sum^{580}_{i=1} \frac{[{\mu}^{obs}(z_{i}) - {\mu}^{th}(z_{i})]^2}{\sigma^2_{i}},\\
B = \sum^{580}_{i=1} \frac{[{\mu}^{obs}(z_i) - {\mu}^{th}(z_{i})]}{\sigma^2_{i}},
\eea
and
\be 
C = \sum^{580}_{i=1} \frac{1}{\sigma^2_{i}}
\ee 
where $\mu^{obs}$ represents the observed distance modulus while $\mu^{th}$ the theoretical one and $\sigma_{i}$ is the uncertainty in the distance modulus.
%%%%%%%%%%%%%%%%%%%%%%%%%%%%%%%%%%%%%%%%%%%%%%%%%%%%%%%%%%%%%%%%%%%%%%%%%%%%
\subsection{BAO/CMB dataset}
Next, we have used baryonic acoustic oscillations (BAO) \cite{bao1,bao2,bao3} and cosmic microwave background (CMB) \cite{cmb} measurements data to obtain the BAO/CMB constraints on the model parameters. The required data points are listed in table \ref{baodata}.
%%%%%%%%%%%%%%%%%%%%%%%%%%%%%%%%%%%%%%%%%%%%%%%%%%%%%%%%%%%%%%%%%%%%%%%%%%%%%%%%
%%%%%%%%%%%%%%%%%%%%%%%%%%%%%%%%%%%%%%%%%%%%%%%%%%%%%%%%%%%%%%%%%%%%%%%
For the BAO/CMB dataset, the details of methodology for obtaining the constraints on model parameters are given in ref. \cite{goistri}. The $\chi^2$ function is defined as 
\be
\chi^2_{BAO/CMB} = X^{T}C^{-1}X 
\ee
with
\be
X=\left( \begin{array}{c}
        \frac{d_A(z_\star)}{D_V(0.106)} - 30.95 \\
        \frac{d_A(z_\star)}{D_V(0.2)} - 17.55 \\
        \frac{d_A(z_\star)}{D_V(0.35)} - 10.11 \\
        \frac{d_A(z_\star)}{D_V(0.44)} - 8.44 \\
        \frac{d_A(z_\star)}{D_V(0.6)} - 6.69 \\
        \frac{d_A(z_\star)}{D_V(0.73)} - 5.45
        \end{array} \right)\,,
\ee
where $d_A(z)=\int_0^z \frac{dz'}{H(z')}$ is the co-moving angular-diameter
distance, $D_V(z)=\left[d_A(z)^2\frac{z}{H(z)}\right]^{\frac{1}{3}}$ is the dilation scale
and $z_\star \approx 1091$ is the decoupling time. Also, the inverse covariance matrix $C^{-1}$ is given by
%%%%%%%%%%%%%%%%%%%%%%%%%%%%%%%%%%%%%%%%%%%%%%%%%%%%%%%%%%%%%%%%%%%%%%%%%%%%%%%%%
\bc
$C^{-1}=\left(
\begin{array}{cccccc}
 0.48435 & -0.101383 & -0.164945 & -0.0305703 & -0.097874 & -0.106738 \\
 -0.101383 & 3.2882 & -2.45497 & -0.0787898 & -0.252254 & -0.2751 \\
 -0.164945 & -2.45499 & 9.55916 & -0.128187 & -0.410404 & -0.447574 \\
 -0.0305703 & -0.0787898 & -0.128187 & 2.78728 & -2.75632 & 1.16437 \\
 -0.097874 & -0.252254 & -0.410404 & -2.75632 & 14.9245 & -7.32441 \\
 -0.106738 & -0.2751 & -0.447574 & 1.16437 & -7.32441 & 14.5022
\end{array}
\right)$
\ec
%%%%%%%%%%%%%%%%%%%%%%%%%%%%%%%%%%%%%%%%%%%%%%%%%%%%%%%%%%%%%%%%%%%%%%%%%%%%%%%%%%%
\par Finally, the total $\chi^{2}$ for these combined observational datasets is given by
\be
\chi^{2}_{total} = \chi^{2}_{H} + \chi^{2}_{SN} +\chi^{2}_{BAO/CMB}
\ee
%%%%%%%%%%%%%%%%%%%%%%%%%%%%%%%%%%%%%%%%%%%%%%%%%%%%%%%
%%%%%%%%%%%%%%%%%%%%%%%%%%%%%%%%%%%%%%%%%%%%%%%%%%%%%%%
%%%%%%%%%%%%%%%%%%%%%%%%%%%%%%%%%%%%%%%%%%%%%%%%%%%%%%%
\begin{table*}
\begin{center}
\begin{tabular}{|c||c|c|c|c|c|c|}
\hline
$z_{BAO}$  & 0.106  & 0.2& 0.35 & 0.44& 0.6& 0.73\\
\hline
$\frac{d_A(z_\star)}{D_V(Z_{BAO})}$ &  $30.95 \pm 1.46$& $17.55 \pm 0.60$
& $10.11 \pm 0.37$ & $8.44 \pm 0.67$ & $6.69 \pm 0.33$ & $5.45 \pm 0.31$
\\
\hline
\end{tabular}
%}
\caption{\it Values of $\frac{d_A(z_\star)}{D_V(Z_{BAO})}$ for different values of $z_{BAO}$.}
\label{baodata}
\end{center}
\end{table*}
%%%%%%%%%%%%%%%%%%%%%%%%%%%%%%%%%%%%%%%%%%%%%%%%%%%%%%%%%%%%%%%%%%%%%%%%
%%%%%%%%%%%%%%%%%%%%%%%%%%%%%%%%%%%%%%%%%%%%%%%%%%%%%%%%%%%%%%%%%%%%%%%%
%%%%%%%%%%%%%%%%%%%%%%%%%%%%%%%%%%%%%%%%%%%%%%%%%%%%%%%%%%%%%%%%%%%%%%%%
%%%%%%%%%%%%%%%%%%%%%%%%%%%%%%%%%%%%%%%%%%%%%%%%%%%%%%%%%%%%%%%%%%%%%%%%%%
%\newpage
%%%%%%%%%%%%%%%%%%%%%%%%%%%%%%%%%%%%%%%%%%%%%%%%%%%%%%%%%%%%%%%%%%%
\section{Results} 
\begin{figure}[!h]
\centerline{\psfig{figure=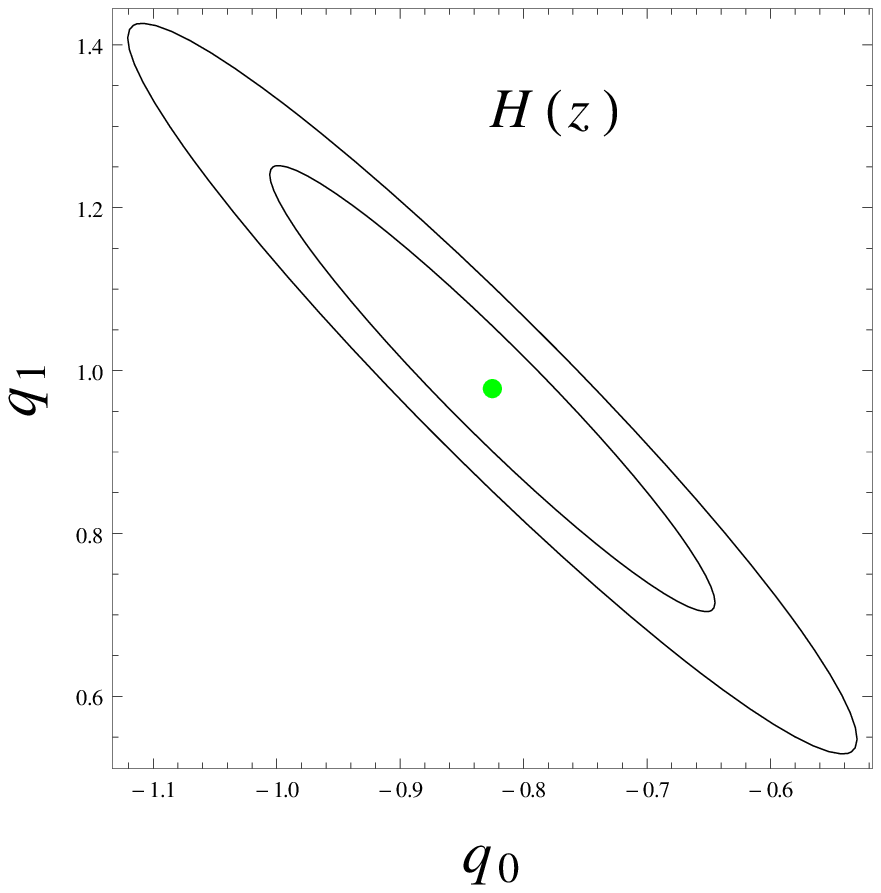,height=60mm,width=60mm}\hspace{3mm} \psfig{figure=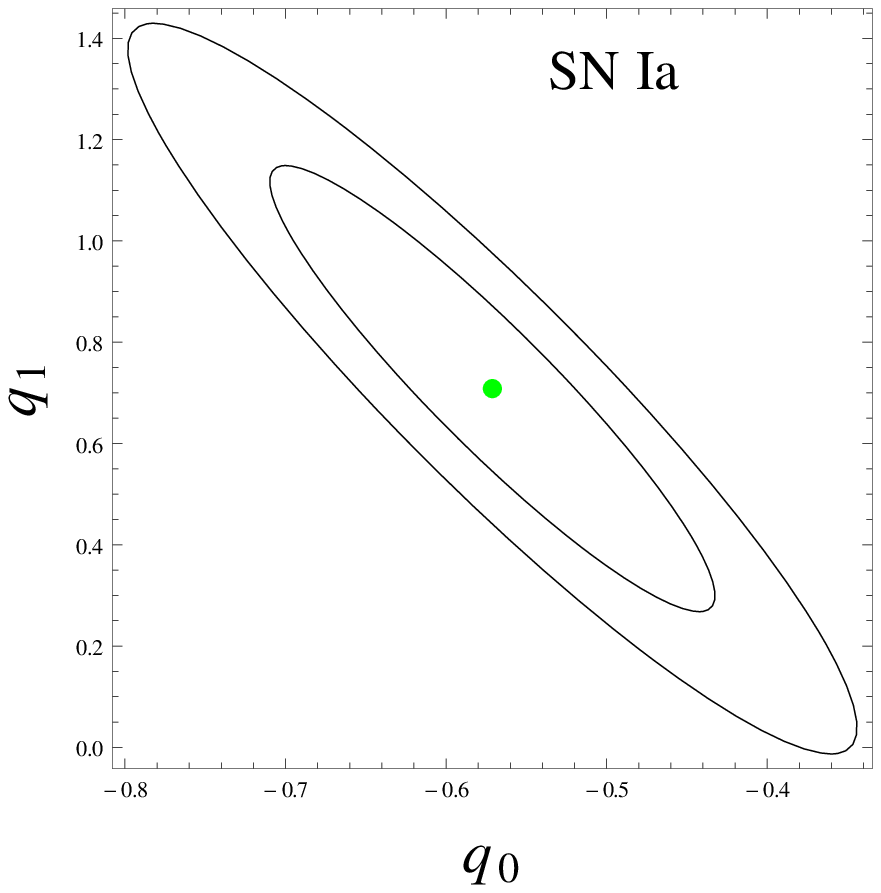,height=60mm,width=60mm}}
\centerline{\psfig{figure=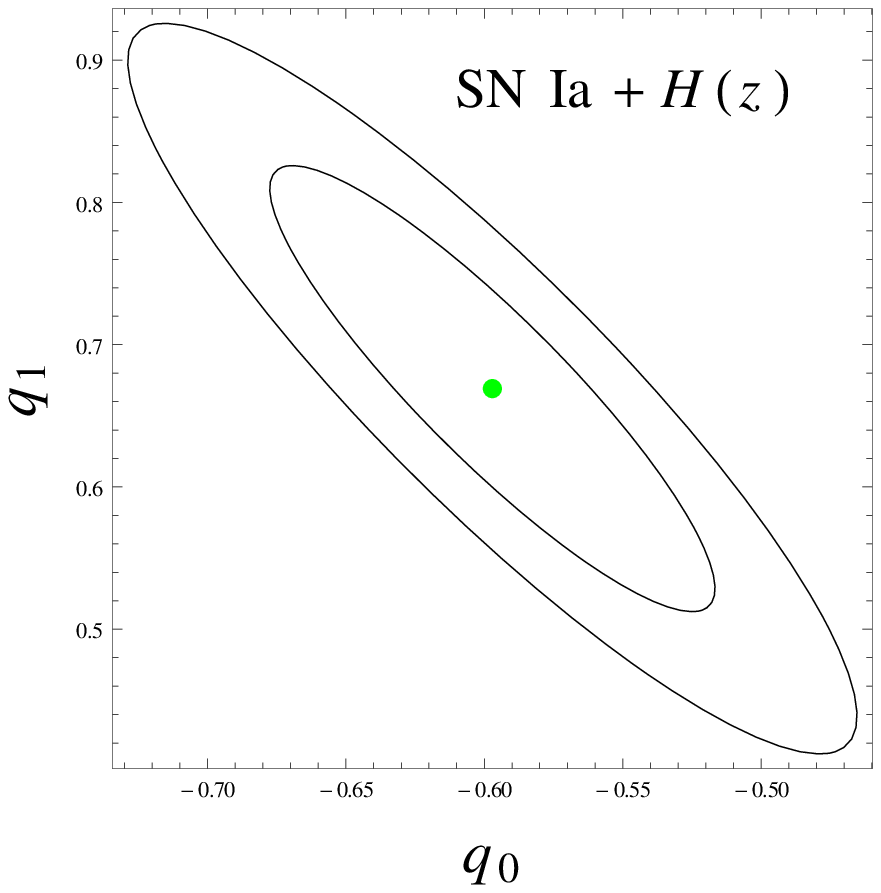,height=60mm,width=60mm}\hspace{3mm} \psfig{figure=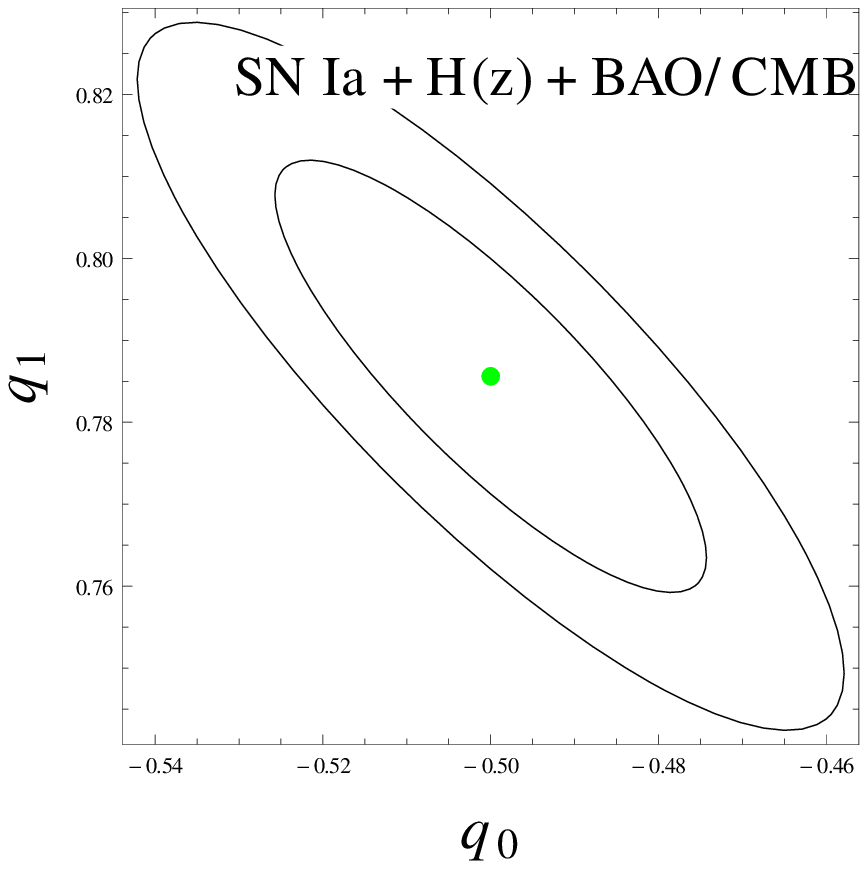,height=60mm,width=60mm}}
\caption{\normalsize{\em Observational constraints on the parameters $q_{0}$ and $q_{1}$ for the parametrization given by equation (\ref{qba}). The contours show the $1\sigma$ and $2\sigma$ confidence level from $H(z)$ (upper left panel), SN Ia (upper right panel), SN Ia $+$ $H(z)$ (lower left panel) and SN Ia $+$ $H(z)$ $+$ BAO/CMB (lower right panel) datasets, for a spatially flat universe. The large dot represents the best fit value of $q_{0}$ and $q_{1}$.}}
\label{figcontour1}
\end{figure}

\begin{figure}[!h]
\centerline{\psfig{figure=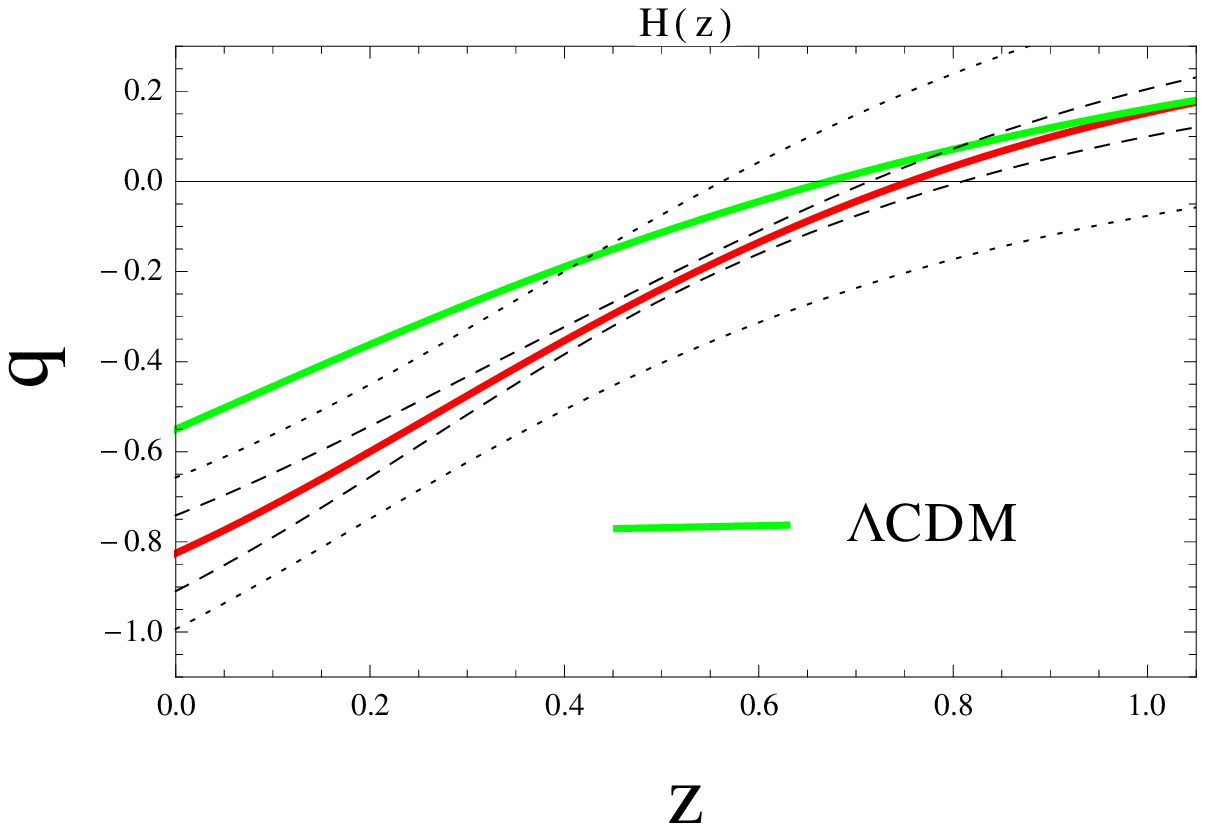,height=50mm,width=60mm}\hspace{3mm} \psfig{figure=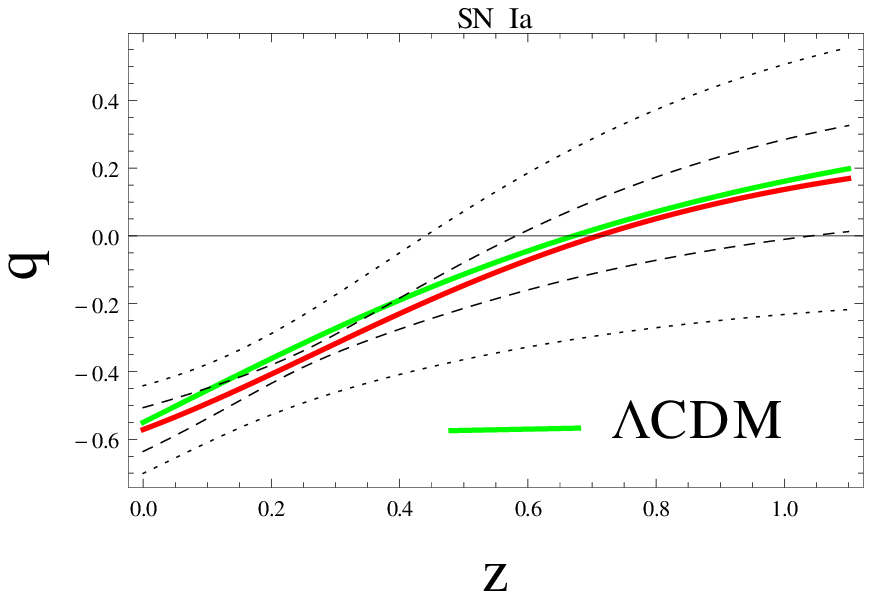,height=50mm,width=60mm}}
\centerline{\psfig{figure=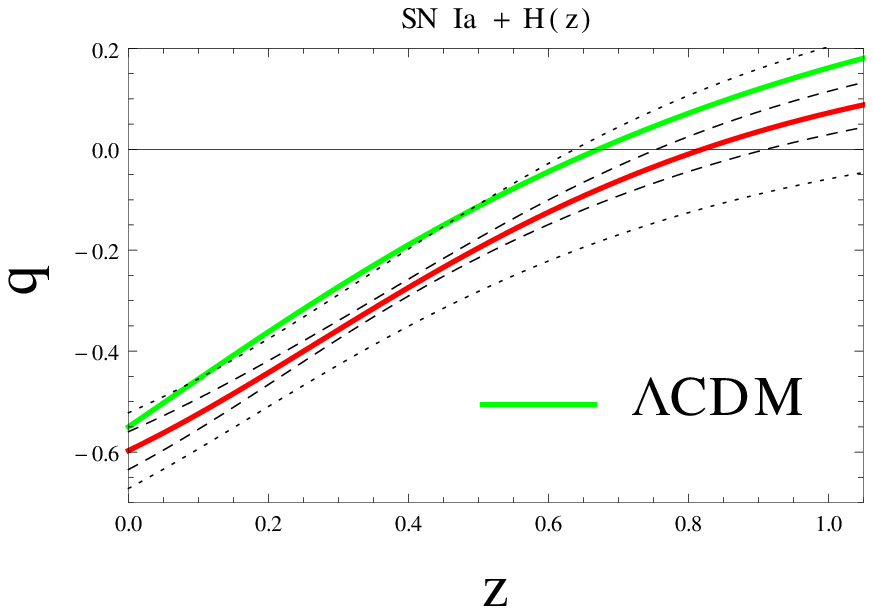,height=50mm,width=60mm}\hspace{3mm} \psfig{figure=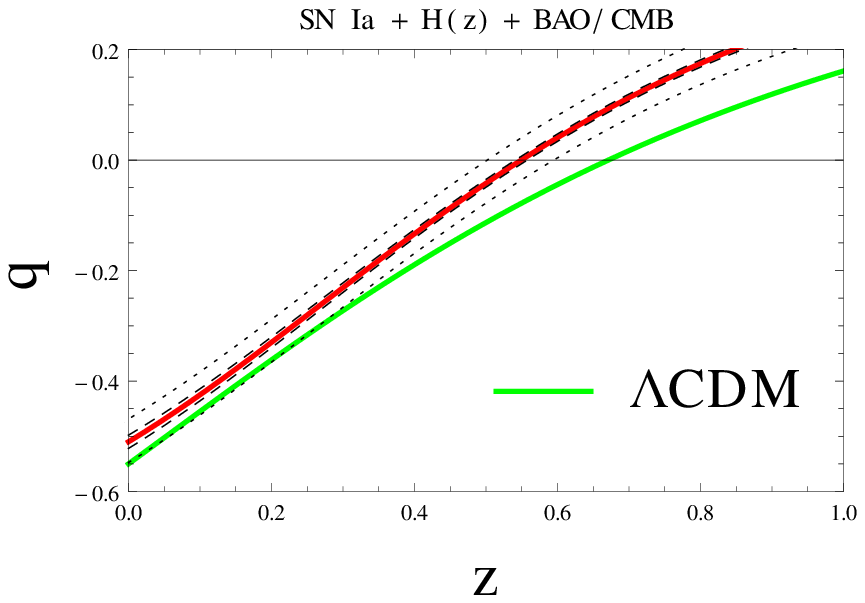,height=50mm,width=60mm}}
\caption{\normalsize{\em Reconstructed deceleration parameter $q$ for this model using various  datasets, as indicated in the each panel. The central thick line (red) represents the best fit curve. The dashed and dotted contour represents the $1\sigma$ and $2\sigma$ confidence level respectively. The green thick line represents the evolution of $q$ in a flat $\Lambda$CDM with $\Omega_{m0}=0.3$ and $\Omega_{\Lambda 0}=0.7$.}}
\label{figcontour2}
\end{figure}

\begin{figure}[!h]
%%%%%%%%%%%%%%%%%%%%%%%%%%%%%%%%%%%%%%%%
\centerline{\psfig{figure=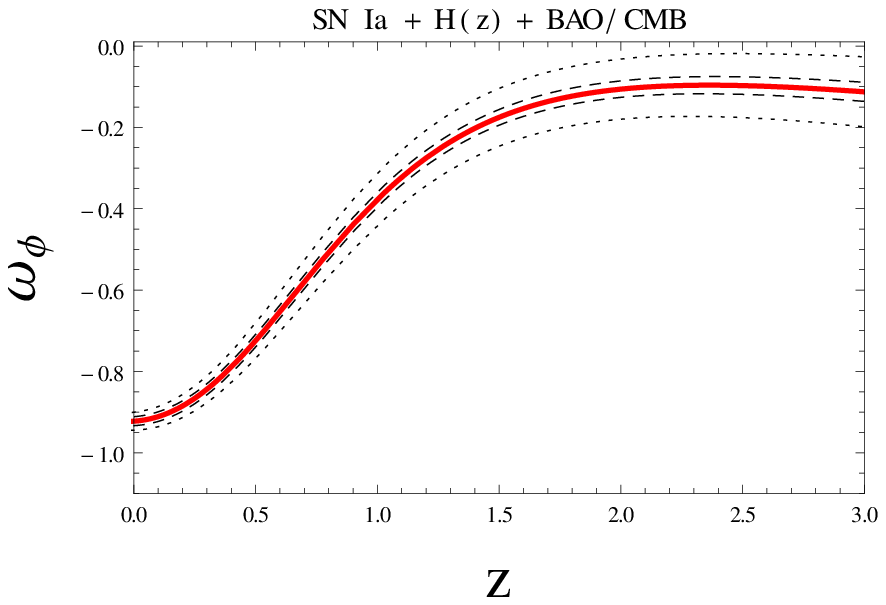,height=50mm,width=60mm}\hspace{2mm} \psfig{figure=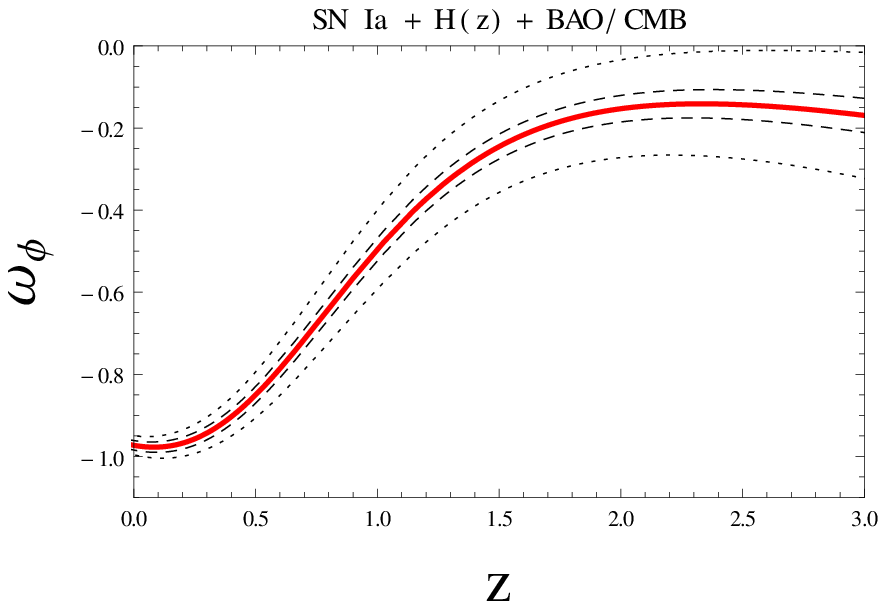,height=50mm,width=60mm}} \centerline{\psfig{figure=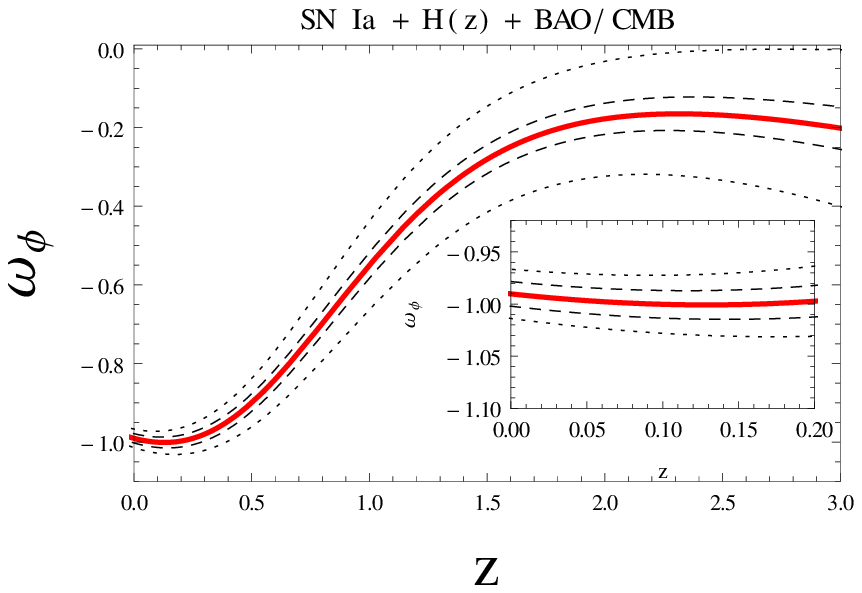,height=50mm,width=65mm}}
\caption{\normalsize{\em The reconstructed equation of state $\omega_{\phi}$ for this model using the combined datasets. The central thick line is for the best fit values of the parameters $q_{0}$ and $q_{1}$ arising from the joint analysis of (SN Ia $+$ $H(z) $+$ BAO/CMB$) dataset. The dashed and dotted contour represents the $1\sigma$ and $2\sigma$ confidence level respectively. In this plot, we have chosen $\Omega_{m0}=0.27$ (upper left panel), $\Omega_{m0}=0.3$ (upper right panel) and $\Omega_{m0}=0.32$ (lower panel).}}
\label{figcontour3}
\end{figure}

In figure \ref{figcontour1} we have shown the $1\sigma$ ($68.3\%$) and $2\sigma$ ($95.4\%)$ confidence contours for $q_{0} - q_{1}$ parameter set for different datasets, as mentioned in each panel.
%%%%%%%%%%%%%%%%%%%%%%%%%%%%%%%%%%%Figures%%%%%%%%%%%%%%%%%%%%%%%%%%%%%%%%%%%%%%%%%%%%
%%%%%%%%%%%%%%%%%%%%%%%%%%%%%%%%%%%%%q0-q1%%%%%%%%%%%%%%%%%%%%%%%%%%%%%%%%%%%%%%%%%%%%
%%%%%%%%%%%%%%%%%%%%%%%%%%%%%%%%%%%%%%%%%%%%%%%%%%%%%%%%%%%%%%%%%%%%%%%%%%%%%%%%%%%%%%% 
The numerical results are displayed in table \ref{bfq}.
%%%%%%%%%%%%%%%%%%%%%%%%%%%%%%%%%%%%%%%%%%%%%%%%%%%%%%%%%%%%%%%%%%%%%%%%%%%%%%%%%%%%%
\begin{table*}
\begin{center}
\begin{tabular}{|c|c|c|c|c|}
\hline
Dataset  & $q_{0}$  & $q_{1}$& ${\chi^{2}_{min}}$ & Constraints on $q_{0},q_{1}$\\
&&&&(within $1\sigma$ C.L.)
\\ 
\hline
%%%%%%%%%%%%%%%%%%%%%%%%%%%%%%%%%%%%%%%%%%%%%%%%%%%%%%%%%%%%%%%%%%%%%%%%%%%%%
H(z) &  $-0.82$& $0.98$& $13.51$ &$-1.01\le q_{0} \le -0.64$\\
&&&&$0.71\le q_{1} \le 1.24$\\
\hline
%%%%%%%%%%%%%%%%%%%%%%%%%%%%%%%%%%%%%%%%%%%%%%%%%%%%%%%%%%%%%%%%%%%%%%%
SN Ia &  $-0.57$& $0.70$& $562.21$ &$-0.70\le q_{0} \le -0.43$\\
&&&&$0.27\le q_{1} \le 1.14$\\
\hline
%%%%%%%%%%%%%%%%%%%%%%%%%%%%%%%%%%%%%%%%%%%%%%%%%%%%%%%%
SN Ia $+$ H(z) &  $-0.59$& $0.67$& $581.18$ &$-0.67\le q_{0} \le -0.52$\\
&&&&$0.51\le q_{1} \le 0.82$\\
\hline
%%%%%%%%%%%%%%%%%%%%%%%%%%%%%%%%%%%%%%%%%%%%%%%%%%%%%%%%55
SN Ia $+$ H(z) + BAO/CMB&  $-0.5$& $0.78$& $628.27$ &$-0.52\le q_{0} \le -0.47$\\
&&&&$0.76\le q_{1} \le 0.81$\\ 
\hline
\end{tabular} 
\caption{\em Best fit values of $q_{0}$ and $q_{1}$ for the present model.}
\label{bfq}
\end{center}
\end{table*}
%%%%%%%%%%%%%%%%%%%%%%%%%%%%%%%%%%%%%%%%%%%%%%%%%%%%%%%%%%%%%%%%%%%%%%%%%%
In figures \ref{figcontour2} and \ref{figcontour3}, we have shown the evolution of the deceleration parameter and equation of state parameter as a function of $z$ respectively. In both the plots, the central thick line is drawn for the best fit values of the parameters. The dashed and dotted contours indicate the $1\sigma$ and $2\sigma$ confidence levels respectively. 
\par In figure \ref{figcontour2}, the green thick line indicates the evolution of $q(z)$ in a standard $\Lambda$CDM model. The four panels are for different datasets as indicated in figure \ref{figcontour2}. It is important to note that when one considers SN Ia $+$ $H(z)$ datasets, the parametrized model is compatible (within 2$\sigma$ limit) with $\Lambda$CDM model, however inclusion of BAO/CMB data makes it incompatible. This indicates that BAO/CMB data puts tighter constraints on model parameters. Inclusion of even higher order terms or some other form of divergenceless parametrizations in $q(z)$ in equation (\ref{qba}) may make better compatibility with $\Lambda$CDM model. However for the present parametrized model it is observed that $q(z)$ is not much compatible with $\Lambda$CDM model for SN Ia $+$ H($z$) $+$ BAO/CMB data.  
%%%%%%%%%%%%%%%%%%%%%%%%%%%%%%%%%%q(z)%%%%%%%%%%%%%%%%%%%%%%%%%%%%%%%%%%%%%%%%%%%%%%%%%
%%%%%%%%%%%%%%%%%%%%%%%%%%%%%%%%%%%%%%%%%%%%%%%%%%%%%%%%%%%%%%%%%
%%%%%%%%%%%%%%%%%%%%%%%%%%%%%%%%%%%%%%%%%%%%%%%%%%%%%%%%%%%%%%%%%%%%%%%%%%%%%%%%%%%%%%%%%%%
\par It is evident form figure \ref{figcontour2} that $q$ undergoes a smooth transition from a decelerating to an accelerating phase of the universe for each dataset. We have also found the observational constraints on the transition redshift $z_{t}$ along with their best fit values  for different datasets and presented them in table \ref{bfz}. 
%%%%%%%%%%%%%%%%%%%%%%%%%%%%%%%%%%Redshift%%%%%%%%%%%%%%%%%%%%%%%%%%%%%%%%%%%%%%%%%%%%%%%
\begin{table*}
\begin{center}
\begin{tabular}{|c|c|c|} 
\hline
Dataset  & $z_{t}$& Constraints on $z_{t}$\\
&(best-fit)&(within $1\sigma$ C.L.)
\\ 
\hline
H(z)& $0.75$ &$0.71\le z_{t} \le 0.81$
\\
\hline
SN Ia & $0.70$ &$0.58\le z_{t} \le 1$
\\
\hline
SN Ia $+$ H(z) & $0.82$ &$0.75\le z_{t} \le 0.91$
\\ 
\hline
SN Ia $+$ H(z) + BAO/CMB & $0.54$ &$0.51\le z_{t} \le 0.57$\\
\hline
\end{tabular}
\end{center}
\caption{\em Best fit values of $z_{t}$ for this model. The constraints on $z_{t}$ are also given. The H($z$), SN Ia and BAO/CMB data has been used in carrying out the $\chi^{2}$ analysis.} 
\label{bfz}
\end{table*}
%%%%%%%%%%%%%%%%%%%%%%%%%%%%%%%%%%%%%%%%%%%%%%%%%%%%%%%%%%%%%%%%%%%%%%%%%%%%%%%%%%%%
The results are found to be consistent with the results obtained independently by several authors (see references \cite{dp3, dp4, dp5, dp7, dp11, zt}), which states that the universe at redshift $z<1$ underwent a phase transition from decelerating to accelerating expansion. However, it is interesting to note that the contours obtained with joint analysis of (SN Ia $+$ $H(z)$ $+$ BAO/CMB) dataset put tighter constraints as compared to the constraints obtained from SN Ia and $H(z)$ data sets (see tables \ref{bfq} $\&$ \ref{bfz}). In figure \ref{figcontour3}, we have plotted $\omega_{\phi}(z)$ for SN Ia $+$ $H(z)$ $+$ BAO/CMB datasets for different values of $\Omega_{m0}$. It is evident from the plots that for $\Omega_{m0}=0.27$ (upper left panel of figure \ref{figcontour3}), the model is not at all compatible with $\Lambda$CDM ($\omega_{\phi}=-1$), but as we increase the value of $\Omega_{m0}$, $\Lambda$CDM model is favoured. However at higher $z$, the situation is completely different. At higher values of $z$, it is seen that the present parametrized model significantly differs from a $\Lambda$CDM model (for which $\omega_{\phi}=-1$). This dynamical behaviour of $\omega_{\phi}$ as evident from equation (\ref{wphi}) is because of the terms involving higher powers of $(1+z)$ and is characteristics of the dynamical nature of dark energy. This obviously may have its implications on the structure formation of the universe. However it is obvious from the plots that at $z\rightarrow 0$, the model approaches $\Lambda$CDM limit. It is always nice to have a $\Lambda$CDM limit for a toy model to be relevant with observations, but as the dynamical nature of dark energy is still unknown, this deviations from $\Lambda$CDM also needs attention.
%%%%%%%%%%%%%%%%%%%%%%%%%%%%%%%%%%%%%%%%%%%%%%%%%%%%%%%%%%%%%%%%%%%%%%%%%%%%%%%%%%%%%% 
%%%%%%%%%%%%%%%%%%%%%%%%%%%%%%%%%%%%%%%%%%%%%%%%%%%%%%%%%%%%%%%%%%%%%%%%%%%%%%%%%%%%%%%%%%%%%%
For the combined dataset, the equation of state parameter $\omega_{\phi}$ is constrained to be $-1.02\le \omega_{\phi} \le -0.96$ at the $2\sigma$ confidence level (with $\Omega_{m0}=0.32$), whereas the best fit value is close to $\omega_{\phi}=-0.99$ at the present epoch (see lower panel of figure \ref{figcontour3}). This result is almost consistent with the recent observational constraints on $\omega_{\phi}$ obtained by Wood-Vasey et al. \cite{wmwvdp} and Davis et al. \cite{tmddp} at low $z$ ($z<0.3$ or so).
\par In ref. \cite{rnnpgprcano} the authors have reconstructed the various parameters of scalar field (for example, potential term, kinetic term and equation of state parameter) from SN Ia and BAO datasets using non-parametric reconstruction method based on a Gaussian Process representation. They have found that the $\Lambda$CDM ($\omega_{\phi}=-1$) model is consistent at $2\sigma$ level for SN Ia dataset and at $1\sigma$ level for BAO dataset. Therefore, such non-parametric techniques may give more robust results which are independent of any ad-hoc parametrization of $q(z)$. Here in our case, as discussed earlier, because of this particular choice of deceleration parameter, $\Lambda$CDM model is not always compatible and may be some different divergence free form of deceleration parameter will satisfy the same.
%%%%%%%%%%%%%%%%%%%%%%%%%%%%%%%%w(z)%%%%%%%%%%%%%%%%%%%%%%%%%%%
%%%%%%%%%%%%%%%%%%%%%%%%%%%%%%%%%%%%%%%%%%%%%%%%%%%%%%%%%%%%%%%%%%%%%%%%%%%%%%%
%%%%%%%%%%%%%%%%%%%%%%%%%%%%%%%%%%%%%%%%%%%%%%%%%%%%%%%%%%%%%%%%%%%%%%%%%%%%%%%%%%%
\par
\begin{figure}[!h]
\centerline{\psfig{figure=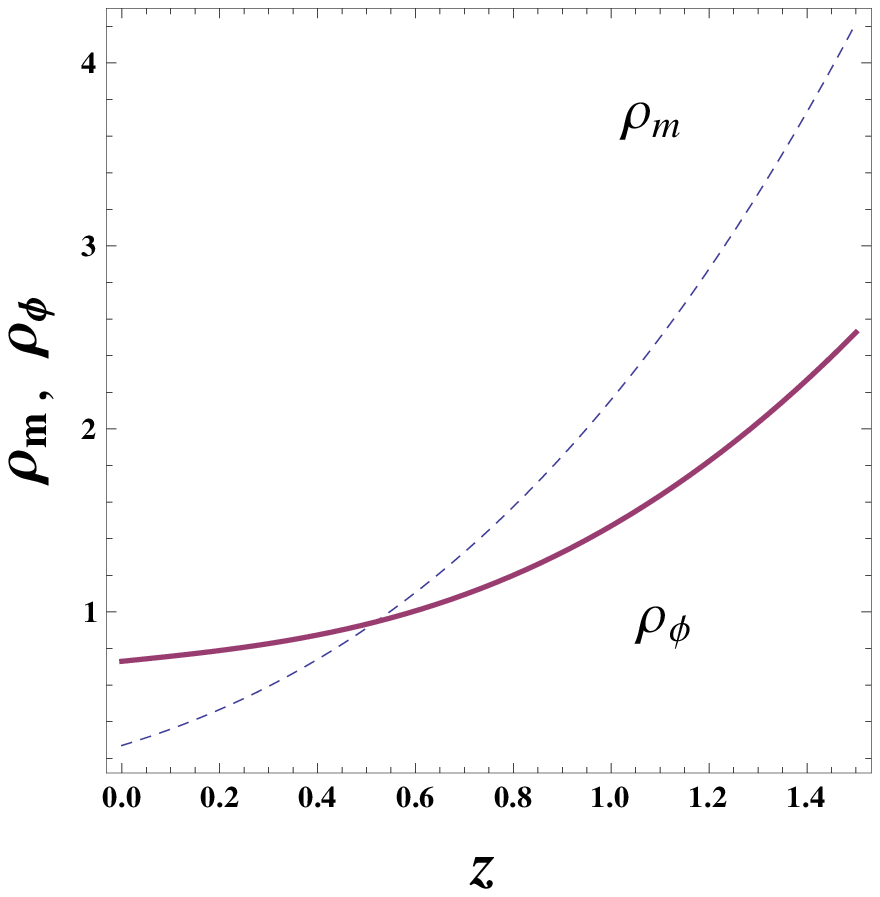,height=50mm,width=60mm}\hspace{3mm} \psfig{figure=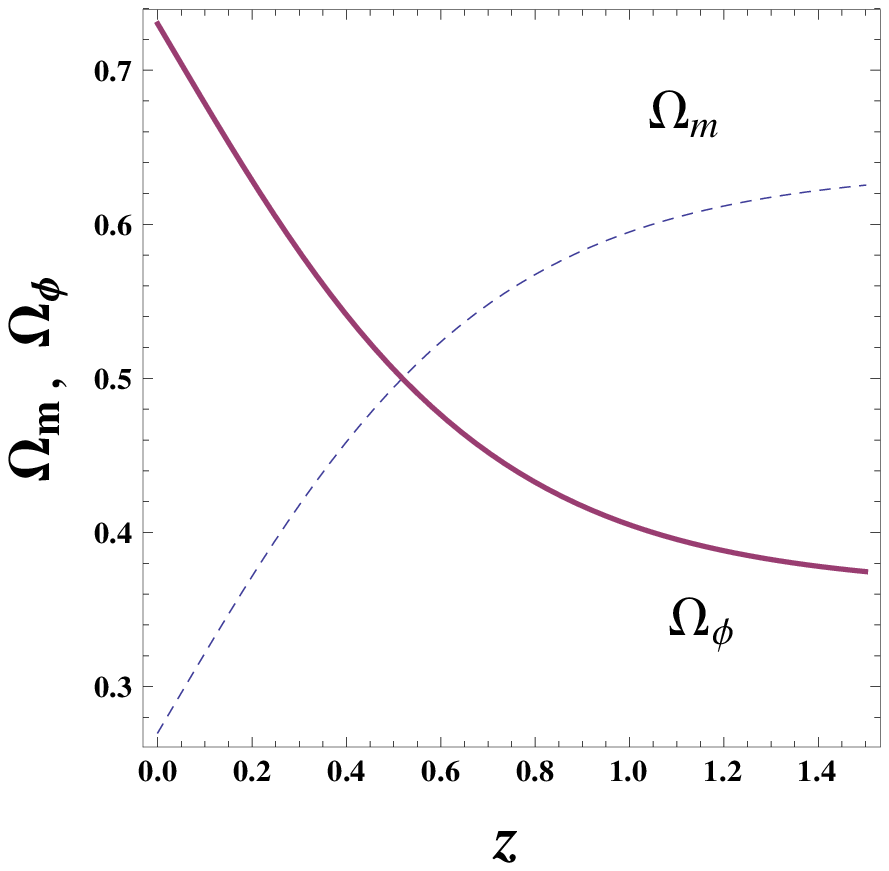,height=50mm,width=60mm}}
\caption{\normalsize{{\em The left and right panels show the evolution of energy densities (in units of critical density) and the behavior of density parameters as a function of redshift $z$ respectively. The plots are for the best fit values of the pair ($q_{0}$, $q_{1}$) arising from the joint analysis of (SN Ia $+$ $H(z) $+$ BAO/CMB$) dataset. In both the plots, we have chosen $\Omega_{m0}=0.27$.} }}
\label{figomegaq}
\end{figure}

 In figure \ref{figomegaq}, we have shown the evolution of the energy densities and the dimensionless density parameters for the scalar field (solid line) and matter field (dashed line). The plot is for the best fit values of the parameters ($q_{0}$, $q_{1}$) for the joint analysis of (SN Ia $+$ $H(z) $+$ BAO/CMB$) dataset.
%%%%%%%%%%%%%%%%%%%%%%%%%%%%%%%%%%%%%%%%%%%%%%%%%%%%%%%%%%%%%%%%%%%%%%
%%%%%%%%%%%%%%%%%%%%%%%%%%%%%%%%%%%%%%%%%%%%%%%%%%%%%%%%%%%%%%%%%%%%%%
%%%%%%%%%%%%%%%%%%%%%%%%%%%%%%%%%%%%%%%%%%%%%%%%%%%%%%%%%%%%%%%%%%%%
\par
\begin{figure}[!h]
\centerline{\psfig{figure=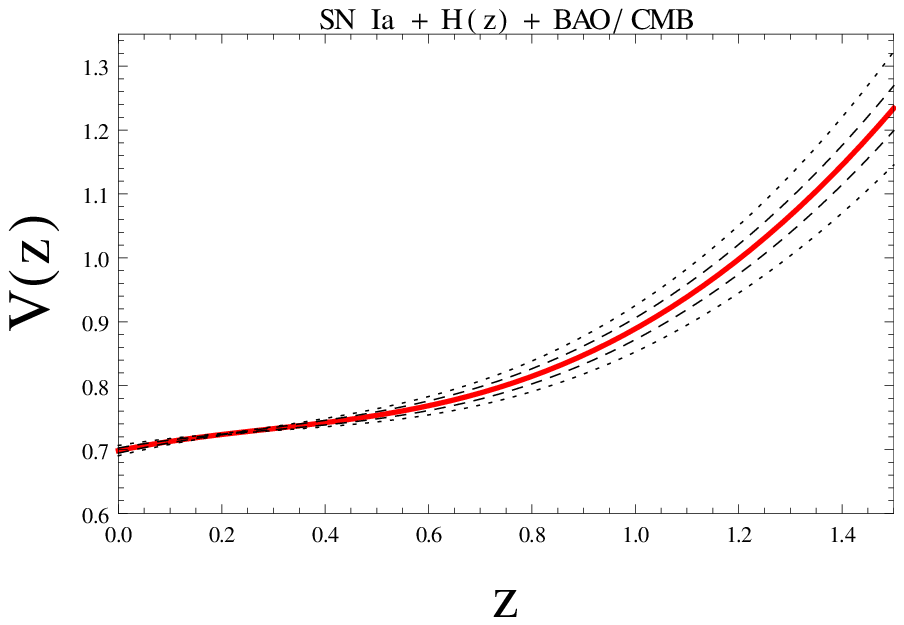,height=50mm,width=60mm}\hspace{3mm} \psfig{figure=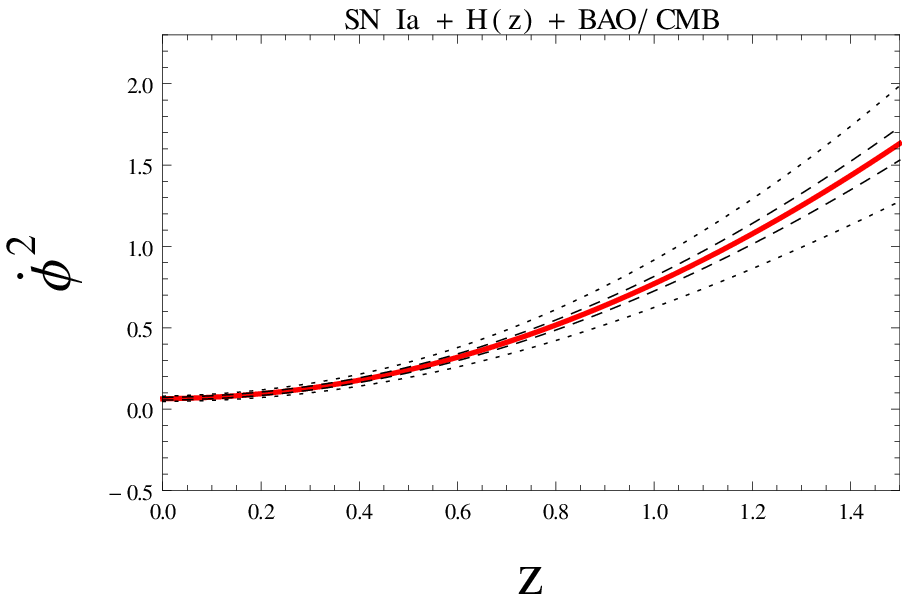,height=50mm,width=60mm}}
\caption{\normalsize{{\em  The reconstructed effective potential $V(z)$ and the kinetic term ${\dot{\phi}}^2$ are shown as a function of redshift $z$ (in units of critical density). The central thick line is for the best fit values of the parameters $q_{0}$ and $q_{1}$ arising from the joint analysis of (SN Ia $+$ $H(z) $+$ BAO/CMB$) dataset. The dashed and dotted contour represents the $1\sigma$ and $2\sigma$ confidence level respectively. In this plot, we have chosen $\Omega_{m0}=0.27$}.}}
\label{figvz}
\end{figure}

 The expression for $\phi(z)$ obtained in equation (\ref{phiz}) is very complicated and it is very difficult to find out the functional dependence of $\phi (z)$. So, we could not express $V$ in terms of $\phi$. However, we have used SN Ia $+$ $H(z) $+$ BAO/CMB$ dataset to reconstruct the form of the effective potential $V(z)$  as well as the kinetic term (${\dot{\phi}}^2(z)$) of the scalar field $\phi$. The reconstructed behavior of $V(z)$ and ${\dot{\phi}}^2(z)$ expressed (in units of critical density) are shown in figure \ref{figvz}.  
%%%%%%%%%%%%%%%%%%%%%%%%%%%%%%%%V(z)%%%%%%%%%%%%%%%%%%%%%%%%%%%%%%%%%%%%%
%%%%%%%%%%%%%%%%%%%%%%%%%%%%%%%%%%%%%%%%%%%%%%%%%%%%%%%%%%%%%%%%%%%%%%
%%%%%%%%%%%%%%%%%%%%%%%%%%%%%%%%V(\phi)%%%%%%%%%%%%%%%%%%%%%%%%%%%%%%%%%%
\begin{figure}[!h]
\centerline{\psfig{figure=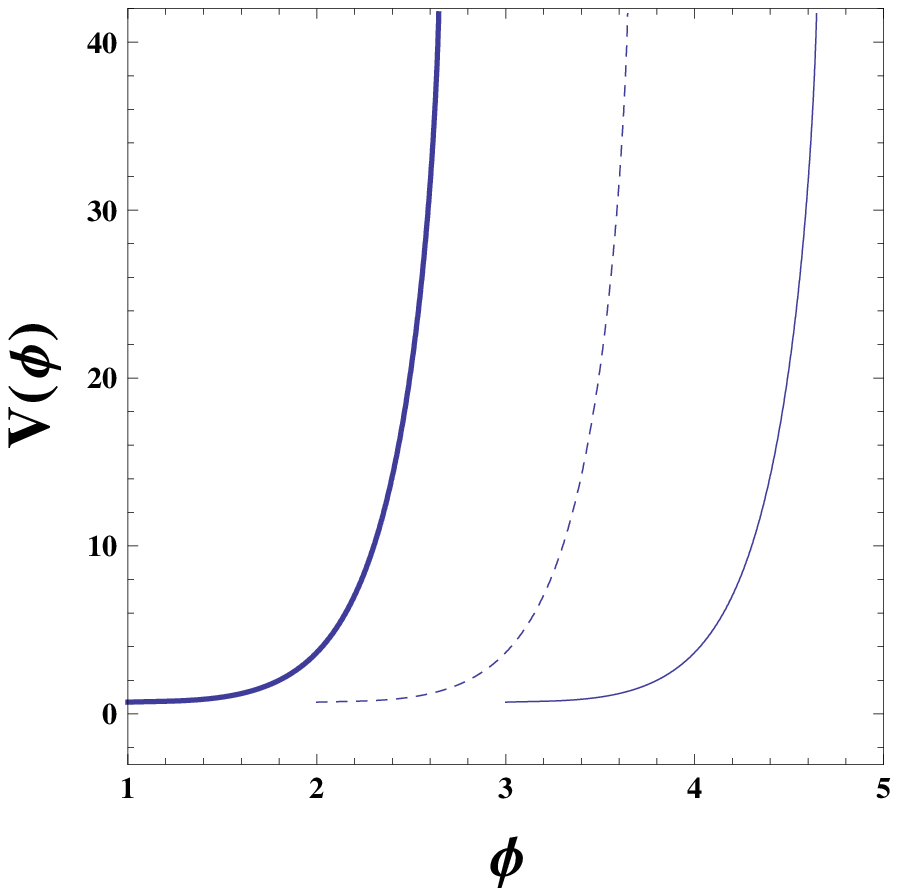,height=40mm,width=40mm}\hspace{2mm} \psfig{figure=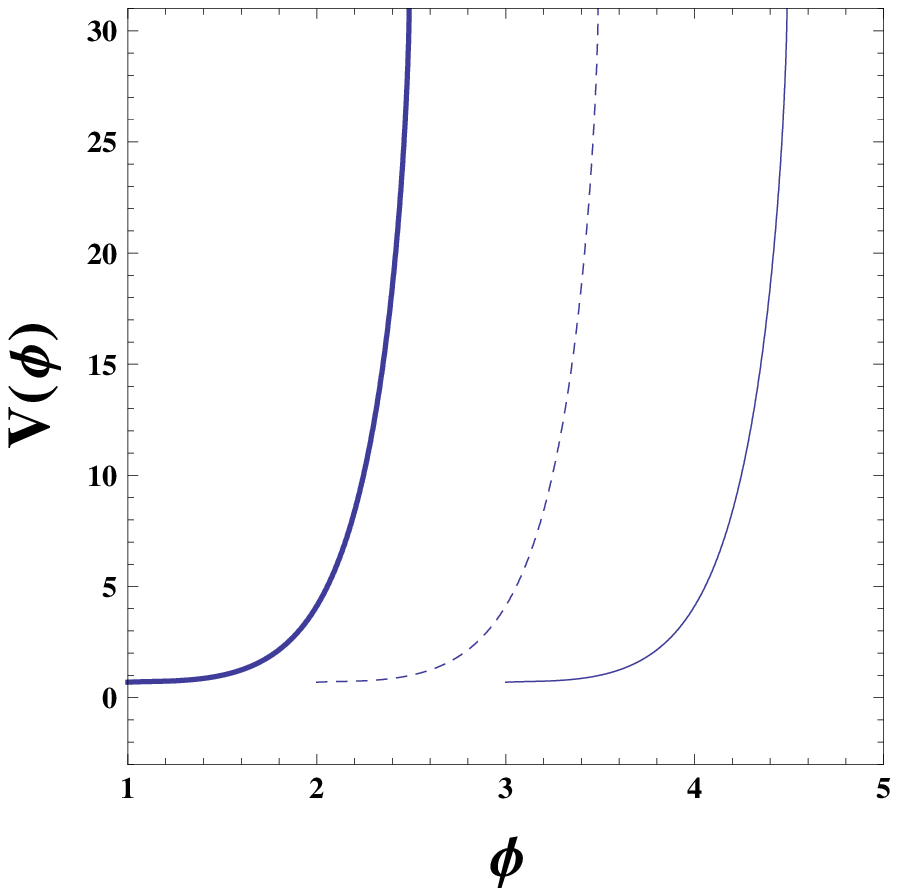,height=40mm,width=40mm}\hspace{2mm} \psfig{figure=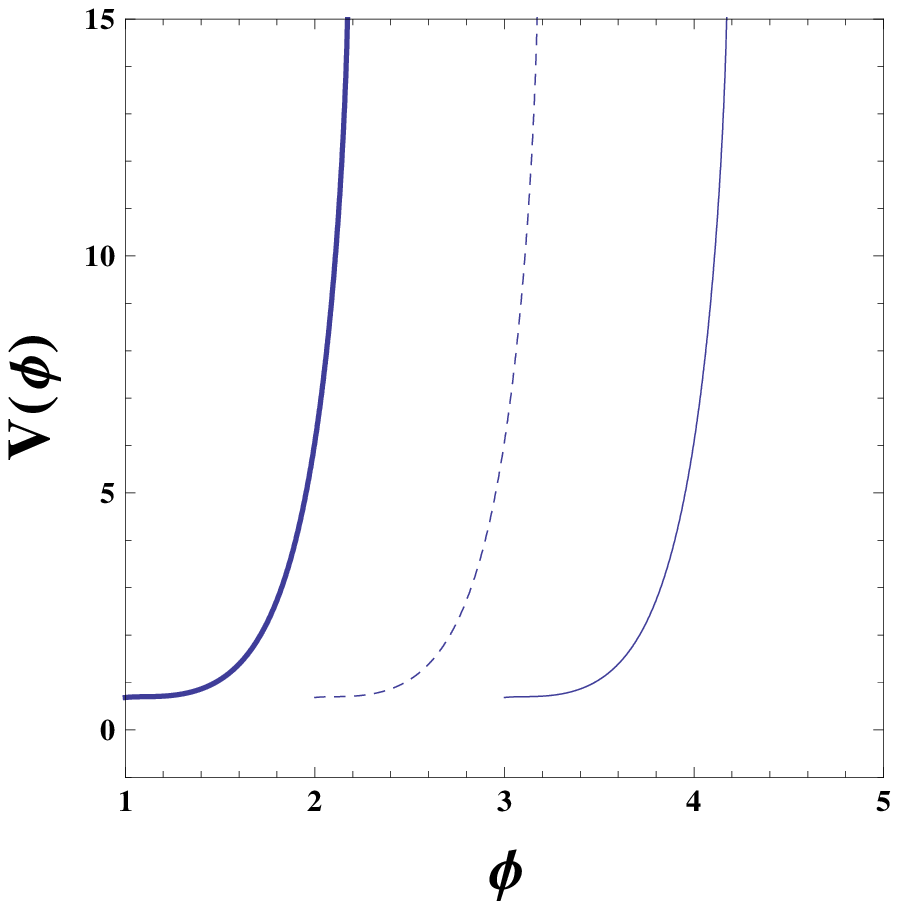,height=40mm,width=40mm}}
\caption{\normalsize{{\em The effective potential $V(\phi)$ (for the potential $\frac{V(z)}{3H^{2}_{0}}$ as given in equation (\ref{vz})) is plotted numerically for the present model with  $q_{0}=-0.5$, $q_{1}=0.78$, $\Omega_{m0}=0.27$ (left panel), $0.28$ (middle panel) and $0.3$ (right panel). The thick, dashed and thin lines are for $\phi_{0}=1, 2, 3$ respectively.}}}
\label{figvphi}
\end{figure}
%%%%%%%%%%%%%%%%%%%%%%%%%%%%%%%%%%%%%%%%%%%%%%%%%%%%%%%%%%%%%%%%%%%%%%%%%%%%%%
From figure \ref{figvz}, we have found that the kinetic term ${\dot{\phi}}^2$ changes very slowly with $z$ as compared to the potential term $V$ at the present epoch. However, we have solved equation (\ref{phiz}) numerically and have plotted the potential $V$ as a function of $\phi$ in figure \ref{figvphi} for different choices of initial conditions (say $\phi_{0}$) and $\Omega_{m0}$. From figure \ref{figvphi}, it is evident that the nature of the behavior of $V$ against $\phi$ remains same by small change in the
value of $\Omega_{m0}$. It is also evident from figure \ref{figvphi} that the trajectory of $V(\phi)$ is independent of the initial conditions. It only shifts the plot horizontally. For $\phi_{0}=2$ and $\Omega_{m0}=0.27$, a simple numerical evaluation shows that the form of the potential can be well approximated as
\be\label{nvphi}
V(\phi)\simeq -V_{1}{\phi}^{2} + V_{2}{\phi}^4 + V_{0}
\ee
where $V_{1}=12.7$, $V_{2}=1.2$ and $V_{0}=33.8$. These values of $V_{1}$, $V_{2}$ and $V_{0}$ has been obtained numerically using the curve fitting technique. It may be noted that we require fine tuning of the parameters $V_{1}$, $V_{2}$ and $V_{0}$ to obtain the exact form of the potential $V(\phi)$. According to the standard model of particle physics, the Higgs potential is given by
\be\label{smvphi}
V(\Phi)=\frac{\lambda}{4}(\Phi^2 - v^2)^{2}=-\frac{\lambda v^2}{2}\Phi^2 + \frac{\lambda}{4}\Phi^4 + \frac{\lambda}{4}v^4
\ee
where $v$ is the vacuum expectation value of the Higgs field $\Phi$ and $\lambda$ represents self-coupling of the Higgs potential. Recently, in the framework of a non-minimally coupled theory, this type of potential have received a lot of attention in the study of the Higgs inflation models \cite{higgs1,higgs2,higgs3}. Comparing equations (\ref{nvphi}) and (\ref{smvphi}), one can say that the canonical scalar field $\phi$ mimics Higgs scalar field for the present model for some particular choice of initial condition. 
%%%%%%%%%%%%%%%%%%%%%%%%%%%%%%%%%%%%%%%%%%%%%%%%%%%%%%%%%%%%%%%%%%%%%%%%%%%%%%%%%%%%
%%%%%%%%%%%%%%%%%%%%%%%%%%%%%%%%%%%%%%%%%%%%%%%%%%%%%%%%%%%%%%%%%%%%%%%%%%%%%%%%%%%%%%
%%%%%%%%%%%%%%%%%%%%%%%%%%%%%%%%%%%%%%%%%%%%%%%%%%%%%%%%%%%%%%%%%%%%%%%%%%%%%%%%
%%%%%%%%%%%%%%%%%%%%%%%%%%%%%%%%%%%%%%%%%%%%%%%%%%%%%%%%%%%%%%%%%%%%%%%%%
\section{Conclusions}
In cosmology, there are a number of proposals to build an acceptable dark energy model to describe the early-time and late-time scenarios of our universe. In literature, the dynamical models of dark energy offer a better framework to investigate the evolution history of the universe. In the framework of a spatially flat FRW universe composed of dark energy and a normal matter field, in this present work, we have made an attempt to construct a viable dark energy model which shows the desired late-time dynamics of the universe. For this purpose, we have considered a simple relation as given in equation (\ref{qba}). Of course, this choice is quite arbitrary and often the chosen parametrization leads to possible biases in the determination of properties of a particular model parameters. Since we are looking for physically viable model of the universe consistent with observations, we make this ansatz in order to close the system of equations. We would like to mention here that the $q$-parametrization adopted in this paper is very simple and valid for the entire redshift range. One the other hand, if we consider more than two terms in the expansion of the unknown function $q(z)$ (see equation (\ref{taylor})), then it would become very difficult to get firm results from data analysis.\\
%%%%%%%%%%%%%%%%%%%%%%%%%%%%%%%%%%%%%%%%%%%%%%%%%%%%%%%%%%%%%%%%%%%%%%%%%%%%%%
\par A remarkable feature of this toy model is that the deceleration parameter $q$ undergoes a smooth transition from a decelerated to an accelerated phase of expansion in the recent past for each dataset (see figure \ref{figcontour2}). As discussed in the previous section, the values of the transition redshift $z_{t}$ are in good agreement with the results obtained by many authors \cite{dp3, dp4, dp5, dp7, dp11, zt}.\\
%%%%%%%%%%%%%%%%%%%%%%%%%%%%%%%%%%%%%%%%%%%%%%%%%%%%%%%%%%%%%%%%%%%%%%%%%%%%%
\par With the $q$-parametrization, we have reconstructed the equation of state $\omega_{\phi}(z)$ and we have shown that the EoS parameter covers a variety of scalar field models (including well-known linear parametrization of $\omega_{\phi}$). In this work, we have found that $q_{0}<0$ and $q_{1}>0$ (within $2\sigma$ confidence level), which shows the expected behavior of $q(z)$ in accordance with the recent observations. Here, $\omega_{\phi}(z)$ does not suffer from the divergency problem like CPL parametrization \cite{cpl,cpl1} at $z=-1$ (see equation (\ref{wphi})) and thus is capable of providing the entire evolution history of the universe.\\
%%%%%%%%%%%%%%%%%%%%%%%%%%%%%%%%%%%%%%%%%%%%%%%%%%%%%%%%%%%%%%%%%%%%%%%%%%%%%%%%%%%%%%%%%%
%%%%%%%%%%%%%%%%%%%%%%%%%%%%%%%%%%%%%%%%%%%%%%%%%%%%%%%%%%%%%%%%%%%%%%%%%%%%%%%%%%%%%%%%
%%%%%%%%%%%%%%%%%%%%%%%%%%%%%%%%%%%%%%%%%%%%%%%%%%%%%%%%%%%%%%%%%%%%%%%%%%%%%%%%%%%
%%%%%%%%%%%%%%%%%%%%%%%%%%%%%%%%%%%%%%%%%%%%%%%%%%%%%%%%%%%%%%%%%%%%%%%%%%%%%%%%%
%%%%%%%%%%%%%%%%%%%%%%%%%%%%%%%%%%%%%%%%%%%%%%%%%%%%%%%%%%%%%%%%%%%%%%%%%%%%%%%%%%
\par We have also plotted the density parameters (also energy densities) of the normal matter and the scalar field as a function of $z$. We have shown that the density parameter of the scalar field composes the major part of the total density parameter ($\Omega_{tot}$ = $\Omega_{m}$ $+$ $\Omega_{\phi}$ ) of the universe at the recent time and the recent transition from a matter dominated to dark energy dominated era occurs at $z\approx 0.53$. This result is in good agreement with the observations.\\
%%%%%%%%%%%%%%%%%%%%%%%%%%%%%%%%%%%%%%%%%%%%%%%%%%%%%%%%%%%%%%%%%%%%%%%%%%
\par With the SN Ia $+$ $H(z)$ $+$ BAO/CMB data, we have also obtained the evolution of the reconstructed potential $V(z)$ for the parametrized deceleration parameter. For the sake of completeness, we have numerically solved for the potential as a function of $\phi$. The relevant potential is found, which mimics a Higgs potential (see equation (\ref{nvphi})) for some particular choice of initial conditions. This, however, requires more detailed and involved analysis because in that case one need not bother about the origin of the scalar field, because the Higgs scalar field has proper theoretical background according to the Standard model of Particle Physics and may provide a new window to probe the nature of dark energy.   
%%%%%%%%%%%%%%%%%%%%%%%%%%%%%%%%%%%%%%%%%%%%%%%%%%%%%%%%%%%%%%%%%%%%%%%%%%%%%%%%%%%%%%%%%%%%%%%%
%%%%%%%%%%%%%%%%%%%%%%%%%%%%%%%%%%%%%%%%%%%%%%%%%%%%%%%%%%%%%%%%%%%%%%%%%%%%%%%%%%%%%%%%%
%%%%%%%%%%%%%%%%%%%%%%%%%%%%%Acknowledgements%%%%%%%%%%%%%%%%%%%%%%%%%%%%%%%%%%%%%
\section{Acknowledgements}One of the authors (AAM) is thankful to Govt. of India for financial support through Maulana Azad National Fellowship. SD wishes to thank IUCAA, Pune for associateship program. The authors are also thankful to the anonymous refree whose valuable comments have improved the quality of this paper.
%%%%%%%%%%%%%%%%%%%%%%%%%%%%%%%%%%%%%%%%%%%%%%%%%%%%%%%%%%%%%%%%%%%%%%%%%% 
%%%%%%%%%%%%%%%%%%%%%%%%%%%%%%%%%%%%%%%%%%%%%%%%%%%%%%%%%%%%%%%%%%%%%%%%%%%%%%%

\end{document}